\documentclass{emulateapj}
\usepackage{url}
\usepackage{epsf}

\shorttitle{iPTF13bvn}
\newcommand{\Ni}{{$^{56}$Ni}}

\shortauthors{Bersten et al.}

\begin{document} 
\title{iPTF13bvn: The First Evidence of a Binary Progenitor for a Type Ib
  Supernova} 
\author{
Melina C. Bersten\altaffilmark{1},
Omar G. Benvenuto\altaffilmark{2,3}, 
Gast\'on Folatelli\altaffilmark{1}, 
Ken'ichi Nomoto\altaffilmark{1,7},
Hanindyo Kuncarayakti\altaffilmark{4,5},
Shubham Srivastav\altaffilmark{6},
G.C. Anupama,\altaffilmark{6}, 
Robert Quimby\altaffilmark{1},
Devendra K. Sahu\altaffilmark{6}}

\affil{\altaffilmark{1} Kavli Institute for the Physics and Mathematics of
  the Universe (WPI), The University of
  Tokyo, 5-1-5 Kashiwanoha, Kashiwa, Chiba 277-8583, Japan}
\affil{\altaffilmark{2} Facultad de Ciencias Astron\'omicas y
  Geof\'{\i}sicas, Universidad Nacional de La Plata, Paseo del Bosque
  S/N, B1900FWA La Plata, Argentina}  
\affil{\altaffilmark{3}  Instituto de Astrof\'\i sica de La Plata (IALP), 
CCT-CONICET-UNLP. Paseo del Bosque S/N (B1900FWA), La Plata, Argentina}  
\affil{\altaffilmark{4} Millennium Institute of Astrophysics, Casilla
  36-D, Santiago, Chile} 
\affil{\altaffilmark{5} Universidad de Chile, Departamento de Astronom\'{\i}a, 
  Casilla 36-D, Santiago, Chile.}
\affil{\altaffilmark{6} Indian Institute of Astrophysics, Koramangala,
  Bangalore 560 034, India}
\affil{\altaffilmark{7} Hamamatsu Professor}
\email{melina.bersten@ipmu.jp}
\submitted{Submitted to AJ on March 28, 2014. Accepted on July 26,
  2014}

\begin{abstract} 
The recent detection in archival HST images of an object at the
the location of supernova (SN) iPTF13bvn may represent the first direct
evidence of the progenitor of a Type~Ib SN. The object's photometry
was found to be compatible with a Wolf-Rayet pre-SN star mass of
$\approx 11$ $M_\odot$. 
However, based on hydrodynamical models we show that the progenitor had a
pre-SN mass of $\approx 3.5$ $M_\odot$ and that it could not be larger
than $\approx 8$ $M_\odot$. We propose an interacting binary system as the SN
progenitor and perform evolutionary calculations that are able to
self-consistently explain the light-curve shape, the absence of
hydrogen, and the pre-SN photometry. We further discuss the range of
allowed binary systems and predict that the remaining companion is
a luminous O-type star of significantly lower flux in the optical
  than the pre-SN object. A future detection of such star may be
    possible and would provide the first robust identification of a
    progenitor system for a Type~Ib SN. 
\end{abstract}


\section{INTRODUCTION} 
\label{sec:intro} 
An important remaining problem in astrophysics is finding the
links between supernovae (SN) and progenitor stars. For
core-collapse SN it is accepted that they arise from massive stars. 
Of particular interest is the origin of hydrogen-deficient SN (Types
Ib and Ic), where the mechanism to remove or deplete the outer
hydrogen envelope is not well determined. The most appealing
alternatives are strong stellar winds in high-mass massive stars ($M
\gtrsim 25 M_\odot$) and mass transfer in close binary systems (see
\citet{langer12} for a recent review). Which is the dominant path for
this type of SN is still unclear and the answer depends on performing
detailed studies of well-observed objects.

A young type Ib SN (He-rich, H-deficient), iPTF13bvn, was
discovered by the Palomar Transient Factory on 2013 June 16
in the nearby galaxy NGC~5806. Using multi-band pre-explosion images 
from HST, a source was identified (within the $2 \sigma$ error box of the SN 
location) as the possible progenitor \citep{cao13}. The 
luminosity and colors of the progenitor candidate are consistent
with some Wolf-Rayet stars \citep{massey06}, one of the proposed 
progenitors of H-deficient SNe. Based on single stellar evolution models,
\citet{groh13b} found that a Wolf-Rayet star with Zero Age Main
Sequence (ZAMS) mass of 31--35 M$_\odot$ was able to reproduce the
observed pre-SN photometry. According to their model, at the moment of
the explosion the star had a mass of 11 $M_\odot$. 

The search for progenitor stars in deep pre-explosion images is a
powerful, direct approach to understand the origin of SNe
and it provides a critical test for stellar evolution models.
Using this technique it was possible to confirm that type
II-P SNe arise from the explosion of red supergiant stars
\citep{smartt09}. But so far no firm progenitor
  identification has been reported for 
H-deficient SNe \citep{yoon12,groh13a,eldridge13}.  
iPTF13bvn may be the first case in its class, thus allowing us for
the first time  to directly  link a SN~Ib with its progenitor.

In most cases, either because the SN is too distant or simply due to 
lacking pre-supernova images, other methods are required to 
infer progenitor properties. One such method is the hydrodynamical
modeling of SN observations. It is a well-known fact that the
morphology of the light curve (LC) is sensitive to the physical
characteristics of the progenitor
\citep[e.g. see][]{nomoto93,blinnikov98}. Therefore, modeling of 
the LC, ideally combined with photospheric velocities or
spectra, provides a useful way to constrain progenitor
properties such as mass and radius, as well as explosion parameters
(explosion energy and production of radioactive material). This
methodology is particularly powerful when combined 
with stellar evolution calculations. A recent example of the
predictability of this technique can be seen in our analysis of the
Type~IIb SN~2011dh \citep{bersten12,benvenuto13}, which allowed us to
provide a self-consistent explanation of the progenitor nature that
was later confirmed \citep{vandyk13,ergon14}. Here we use the same
approach to address the problem of the progenitor of iPTF13bvn.

The observational material used in this work is briefly described in
\S~\ref{sec:data}. In \S~\ref{sec:hydro} we present our hydrodynamical
modeling of iPTF13bvn with focus on determining the mass and
size of the progenitor. In \S~\ref{sec:binary} we further analyze the
origin of iPTF13bvn and of the pre-explosion object, and we propose an
interacting binary progenitor. The range of allowed binary models is
discussed in \S~\ref{sec:companion}, where we also predict the
feasibility of detecting of the companion star in future
observations. In \S~\ref{sec:conclusion} we present the main
conclusions of this work.

\section{OBSERVATIONAL DATA}
\label{sec:data}
We computed the observed bolometric LC of iPTF13bvn based on 
$UBVRI$ photometry obtained by \citet{shubham14}, and adopting bolometric 
corrections derived for core-collapse SNe by \citet{lyman14}. The
bolometric corrections were based on $(B-V)$ colors, although other
calibrations produced compatible results. To correct colors and
magnitudes to intrinsic values, we adopted a Milky-Way reddening of
$E(B-V)_{\mathrm{MW}} = 0.0447$ mag from the NASA/IPAC Extragalactic
Database (NED) \citep{schlafly11}. For estimating reddening in the
host galaxy we compared the Milky-Way reddening-corrected $(B-V)$
colors with an intrinsic-color law derived from a sample of SE SNe
observed by the Carnegie Supernova Project \citep{stritzinger}. The
average difference with the intrinsic colors, given in the interval
between $B$-band maximum light and 20 days after, yielded a color
excess of $E(B-V)_{\mathrm{host}}=0.17 \pm 0.03$ mag. We further checked this
color excess with measurements of the equivalent width of
the \ion{Na}{1}~D lines in the spectra of iPTF13bvn. With a measured
$EW_{\mathrm{Na\,I\,D}}=0.5$ \AA\ and adopting the relations between
$EW_{\mathrm{Na\,I\,D}}$ and color excess given by \citet{turatto03},
we obtained $E(B-V)=0.07$ mag or $0.22$ mag, depending on the adopted
relation. These values can be interpreted as an interval of valid
solutions given the dispersion of the data used to derive the
relations. We consider the estimate obtained from the $(B-V)$ colors
to be more robust and the result to be confirmed by the \ion{Na}{1}~D
measurements. We thus adopted $E(B-V)_{\mathrm{host}}=0.17 \pm 0.03$
mag to correct the SN observations. We employed the same value for the
pre-SN observations under the assumption that the pre-explosion object
was affected by the same amount of reddening as the SN. This may be
an underestimate of the pre-SN extinction if the explosion destroyed
part of the dust that was obscuring the progenitor. Finally, to obtain
luminosities we used the average distance to NGC~5806 of $25.54 \pm
2.44$ Mpc, as provided by NED. A representative uncertainty of $0.1$
  dex in luminosity is assumed throughout this paper. This value was
  estimated from the uncertainties in distance, extinction
  and bolometric corrections, summed in quadrature. The uncertainty is
  dominated by the systematic error in distance and is almost
  constant in time. The LC points may move vertically by the indicated
amount, but the shape of the LC would not change significantly. 

In this paper we assumed an explosion time ($t_{\mathrm{exp}}$) of
JD\,$=2456459.24$, i.e., the mid-point between the last non-detection
and the discovery date \citep[see][]{cao13}. We considered an
  uncertainty of $0.9$ day in the explosion time, based on the
  interval between the last non-detection and the discovery. 
We emphasize that this uncertainty on $t_{\mathrm{exp}}$ is
robust due to the stringent detection limit and existing early
observations in $R$ band (see Section~\ref{sec:hydro}). Our
bolometric LC shows a rise time to 
maximum of $\Delta t \approx 16$ days, i.e., near the lower end of the
typical range for normal stripped-envelope SNe (SE SNe)
\citep{Richardson06}. The peak luminosity is $L_{\mathrm{peak}}\approx
1.8 \times 10^{42}$ erg. 

In addition to the bolometric luminosities, in \S~\ref{sec:hydro} we
tested our model's photospheric velocity against expansion velocities
measured from the spectra. For that purpose, we adopted the
\ion{Fe}{2} expansion velocities from \citet{shubham14}. Part of the analysis in
Section~\ref{sec:hydro} also aimed at determining the radius of the
progenitor star based on modeling the early-time light curve. For this
purpose we used the $R$-band light curve presented by \citep{cao13}
because it provided the earliest coverage, starting at less than one
day after discovery. 

\section{HYDRODYNAMICAL MODELING} 
\label{sec:hydro}
We performed a set of explosion models using our one-dimensional LTE
radiation hydrodynamics code \citep{bersten11}. Since we were
investigating a H-free SN, as initial structures we adopted helium
stars of different masses. Specifically, we tested
models with $3.3$ $M_\odot$ (HE3.3), 4 $M_\odot$ (HE4), 5 $M_\odot$
(HE5), and 8 $M_\odot$ (HE8), which correspond to 
main-sequence masses of 12, 15, 18, and 25 $M_\odot$, respectively
\citep{NH88}. All these initial configurations have compact
structures with radii of $R < 3\, R_\odot$. More details about the initial
models can be found in \citep{bersten12}.

\begin{figure}[htpb]
\begin{center}
\includegraphics[scale=.29,angle=-90]{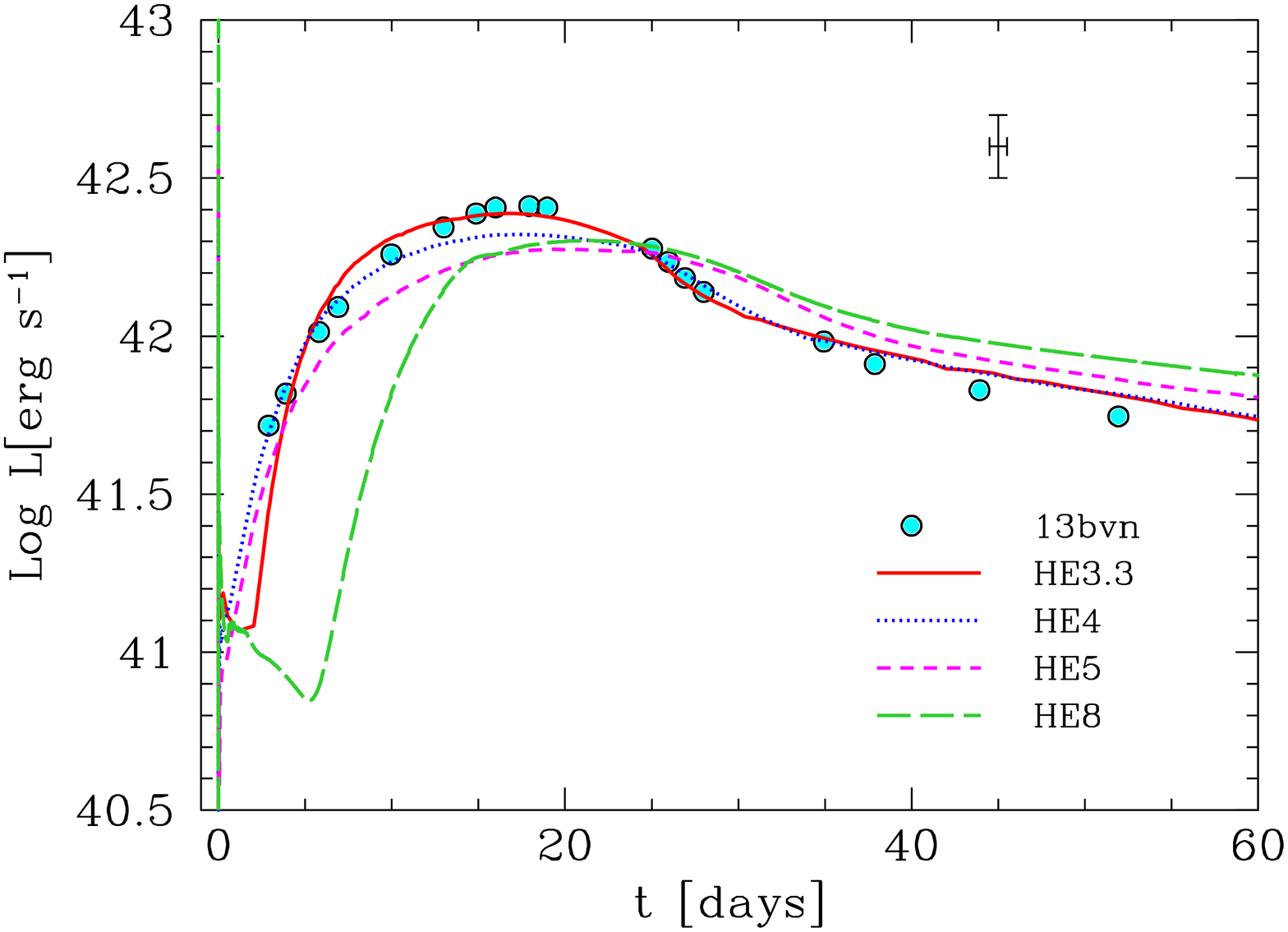}\\
\includegraphics[scale=.29,angle=-90]{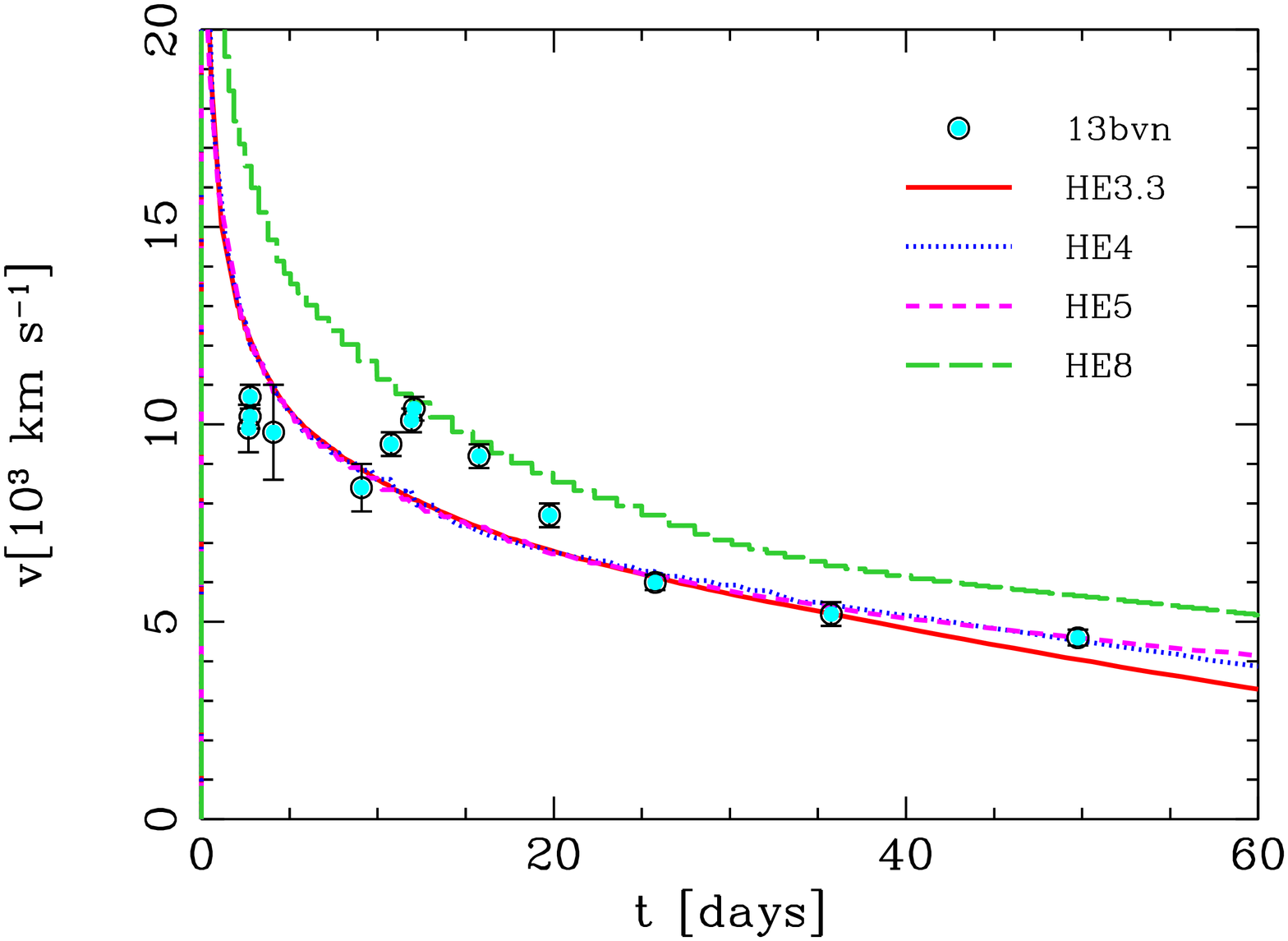} \\
\caption{\label{fig:optimal} Hydrodynamical modeling of
  iPTF13bvn. Bolometric light curve ({\em upper panel\,})
  and photospheric velocity evolution ({\em lower panel\,}) 
 are compared with observations (dots). Models with different
masses are shown with different line types and colors. HE3.3 and HE4
give a good representation of the observations but a slightly more massive
object, HE5, provides a worse comparison. A model with 8 $M_\odot$
 (HE8) is clearly not acceptable. The error bars at the top of
  the figure indicate the nearly constant uncertainty in luminosity
  and the adopted uncertainty in the explosion time (see
  \S~\ref{sec:data})}.   
\end{center} 
\end{figure}

SN parameters such as explosion energy ($E$), ejected mass
($M_{\mathrm{ej}}$), and the mass ($M_{\mathrm{Ni}}$) and
distribution of \Ni\, synthesized during the 
explosion were derived chiefly by analyzing the main peak of the
bolometric LC and the photospheric velocity ($v_{\mathrm{ph}}$)
evolution. At first order, the evolution of $v_{\mathrm{ph}}$ is
not sensitive to $M_{\mathrm{Ni}}$ and its distribution. Therefore,
using the observed expansion velocity, a constraint on $E$ for each
initial model (He mass) could be derived, and then $M_{\mathrm{Ni}}$
and mixing were adjusted to reproduce the bolometric
LC. Figure~\ref{fig:optimal} shows our 
hydrodynamical modeling for iPTF13bvn. While HE3.3 and HE4 can produce
reasonably well the observations (LC and velocities), a slightly more
massive model, HE5, already shows a worse agreement. Note that the
uncertainty in the observed luminosity is nearly systematic and 
thus the LC could shift vertically but without changing its
shape. The main parameters affected by this shift are
the \Ni\ mass and its distribution \citep[see Figures~4 and 5
  of][]{bersten12}. Thus the predicted LC for model HE5 is too wide
compared with the observations, regardless of any overall shift of the
data points.
Both HE3.3 and HE4 with similar SN parameters appear to be equally
plausible. We thus consider our best solution to be that with
intermediate parameters between those models. That means that
iPTF13bvn was produced by the explosion of a low-mass He star of
$\approx 3.5$ $M_\odot$, with an ejected mass of $M_{\mathrm{ej}}
\approx 2.3$ M$_\odot$\footnote{$M_{\mathrm{ej}}= M_{\mathrm{tot}} -
  M_{\mathrm{cut}}$, where $M_{\mathrm{tot}}$ is the total mass of the
  He star and $M_{\mathrm{cut}}$ is the mass assumed to to form a
  compact remnant}, an explosion energy of $E= 7 \times 
10^{50}$ erg and a \Ni\ yield of $M_{\mathrm{Ni}} \approx
0.1$ $M_\odot$. These parameters indicate iPTF13bvn was
a low-energy event with normal nickel production.
 Also, we found that
a quite strong mixing (extending to $\approx 96$\% of the initial mass) was
required in all the calculations to reproduce the rise time of the LC
which in turn is very well known due to the strong constraint on
the explosion time \citep[$t_{\mathrm{exp}}$;][]{cao13}. 

\begin{figure}[htpb]
 \begin{center}
\includegraphics[angle=-90,width=9.cm]{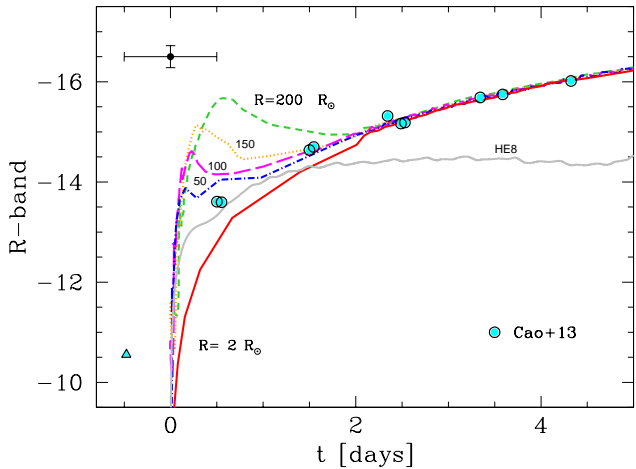}
\caption{\label{fig:rband} Early $R$-band light curve of iPTF13bvn (dots)
compared with models of different progenitor radii (lines). The labels
next to the curves indicate the radius in $R_\odot$.
The radius variation is accomplished by attaching essentially mass-less
($<0.01$ $M_\odot$) envelopes to the compact He-star model HE4 (see
\S~\ref{sec:hydro}). In spite of the good constraint on
$t_{\mathrm{exp}}$, its uncertainty (black line) is still too large
to distinguish between a compact star (of a few $R_\odot$) and
relatively relatively extended structure (of $R \lesssim 150\, R_\odot$). 
A better constraint on $t_{\mathrm{exp}}$ or a higher cadence of the
observations is required to capture the short-duration emission
feature produced by relatively extended progenitors. The error
  bars indicate the size of the uncertainty in magnitude (dominated by
  the distance uncertainty) and in the explosion time (see
  \S~\ref{sec:data}). Model HE8 is included to show that a it would
  not be compatible with a shift in the explosion date of more than
  $\approx$ 1 day.}
\end{center}
\end{figure}

The low progenitor mass suggested by our modeling
is in clear contradiction with the range of masses allowed for Wolf-Rayet
stars \citep[e.g., see][]{heger03,groh13a}, and thus our results
are in disagreement with those of \citet{groh13b}. Specifically, in
Figure~1 we also show the case of a He star with 8 $M_\odot$ for
  an explosion energy of $3 \times 10^{51}$ erg. We
  show a compromise solution for this model that produces too high
  velocities and too wide LC. If we adopted a lower value of the
  explosion energy in order to better fit $v_{\mathrm{ph}}$, then the
  LC would become even fainter and wider, in worse agreement with the
  observations. Note also that although the 
  HE8 model shown in Figure~\ref{fig:optimal} could be made more
  compatible with the observations by shifting the explosion date to
  about one week earlier, that would be incompatible with the
  non-detection limit in $R$ band that is shown in
  Figure~\ref{fig:rband}. Even considering all the uncertainties
related with the model hypotheses and with the 
observations, we can firmly rule out models with He core mass $\gtrsim
8$ $M_\odot$ as progenitors of iPTF13bvn.

The compactness of the progenitor can be explored if the SN is
observed early enough, before its emission becomes powered by \Ni\,
decay. The shape of the LC in this early phase is given by the
progenitor size and, to some extent, by the degree of \Ni\, mixing.
The good constraint on $t_{\mathrm{exp}}$ and the absence of an
initial peak in the early LC of iPTF13bvn can lead to the naive 
conclusion that the progenitor should have been a compact star of a
few solar radii in size. We have tested this by attaching thin
He-rich envelopes of different radii to our HE4 model, as described in
\citet{bersten12}\footnote{Note that low-mass HE stars can experience
  an expansion of their outer envelope, as shown by \citet{yoon10} (see also
\S~\ref{sec:binary})}. We derive $R$-band photometry from the models by 
assuming a black-body spectral energy
distribution. Figure~\ref{fig:rband} shows the early $R$-band LC 
compared with our models. From the comparison, it is clear that models with
relatively extended structures, $R\lesssim 150\; R_\odot$, cannot be
ruled out considering the uncertainty in $t_{\mathrm{exp}}$ ($\approx 0.9$
day). Therefore the progenitor of iPTF13bvn is not necessarily a
compact star. Interestingly, our modeling suggests that in order to
to capture the differences in LC shapes and thus to discriminate
between compact and relatively extended progenitors, it is necessary
to obtain several observations during the same night. 

\section{BINARY PROGENITOR}
\label{sec:binary}
The mass we derived from hydrodynamical modeling is difficult to reconcile 
with the idea of a single progenitor for the Type~Ib SN
iPTF13bvn. In order to remove the hydrogen envelope, a single star
should be massive enough ($M_{\mathrm{ZAMS}} \gtrsim 25$ $M_\odot$) to
produce strong stellar winds that are able to strip the envelope during its
evolution \citep[see, e.g.,][]{langer12}. However, in that case the
resulting helium star mass would be too large ($M_{\mathrm{He}}
\gtrsim 8$ $M_\odot$) to account for the observed SN LC, as shown in
\S~\ref{sec:hydro}. Alternatively, the path to the explosion of a
low-mass He star is naturally provided by interacting binaries. 
 Although \citet{cao13} derived a mass-loss
  rate that is compatible with a massive Wolf-Rayet star based on
  radio observations, this result depends strongly on the assumed
  wind velocity among other parameters that are not completely
  known. For a low-mass helium star a lower mass-loss rate is expected
  but the wind properties are uncertain, therefore the radio
  observations cannot be used to validate or reject this 
  possibility.
 
The question is if there are binary configurations capable of
simultaneously reproducing the SN properties 
and the pre-explosion photometry. 

\begin{figure}[htpb]
\begin{center}
\includegraphics[angle=-90,scale=.33]{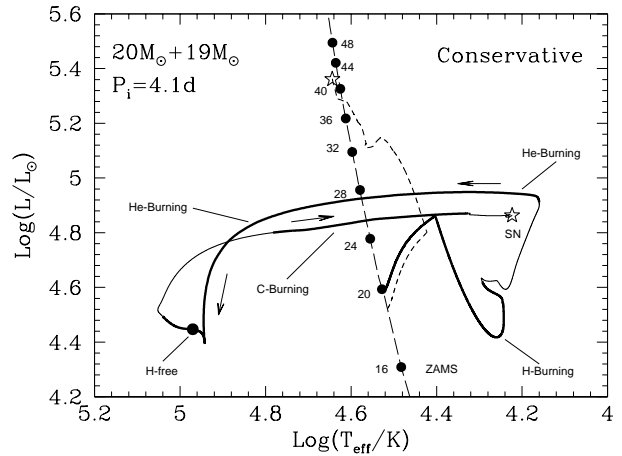}   
\caption{\label{fig:binary} Evolutionary tracks of the binary
  components of the progenitor of iPTF13bvn for a proposed system
  with initial masses of 20 $M_\odot$ and 19 $M_\odot$ and an initial
  orbital period of $4.1$ days. The solid line indicates the track of
  the primary (donor) star (arrows show the evolutionary
  progress). The short-dashed line shows the evolution of the
  secondary (accretor) star. Fully conservative accretion ($\beta=1$)
  is assumed. The star symbols show the location of
  both components at the moment of explosion of the primary
  star. Thick portions of the primary's track indicate the phases of 
  nuclear burning at the stellar core.
 The long-dashed line shows the locus of the ZAMS, with dots showing
 different stellar masses (labels in units of $M_\odot$).}  
\end{center} 
\end{figure}

To address this question we used the binary stellar evolutionary code
employed in our analysis of SN~2011dh \citep{benvenuto13} which is a
recent update of the code presented by \citet{benvenuto03}. For the
present problem, we considered non-rotating stars. At 
mass transfer conditions, the code solves the donor star structure,
the mass transfer rate and the orbital evolution simultaneously
in a fully implicit way, providing numerical stability
\citep[see][]{benvenuto03}. Figure~\ref{fig:binary} 
shows the evolutionary tracks in the H-R diagram
for a system composed by a donor
(primary) star of 20 $M_{\odot}$ and an accretor (secondary) star
of 19 $M_{\odot}$ on a circular orbit with initial period of $4.1$
days. We further assumed conservative mass transfer,
i.e., that all the mass that is transferred is accreted onto the
secondary (parameterized by the accretion efficiency, $\beta$=1). 
The choice of the initial masses and orbital period of the system was 
guided to account for the composition of the exploding star, the
LC of the supernova, and the available photometric observations prior
to the explosion. Such configuration is by no means unique but it
serves to demonstrate the feasibility of the binary progenitor scenario.
We will explore the range of possible progenitor systems in greater
detail in \S~\ref{sec:companion}.   

The system undergoes class A mass transfer---i.e., it experiences the first
Roche-lobe overflow (RLOF) when it is still in the core hydrogen
burning stage---, it detaches shortly after core hydrogen exhaustion and, 
as consequence of the formation of a shell burning hydrogen, suffers a
second RLOF episode until core helium ignition. After detachment, the
donor star evolves bluewards up to very high effective
temperatures. At these stages the star undergoes wind mass loss that
removes the hydrogen-rich outer layers. The mass loss has been
  treated as in \citet{yoon10}, setting the correction factor for
  Wolf-Rayet winds defined in their Equation~(1), $f_{\mathrm{WR}}$,
  to $2.5$.

After helium core exhaustion the donor star evolves redwards
reaching the final pre-SN structure. We stopped the calculation at
oxygen core exhaustion, assuming that no significant displacement in the
H-R diagram occurs until core collapse. In the meantime, the
companion star accretes hydrogen-rich material, which makes it 
to brighten. As the accretion rate is rather low, this star moves
upwards in the H-R diagram without swelling appreciably. In this way
the system avoids undergoing a common envelope episode. 

\begin{figure}[htpb]
\begin{center}
\includegraphics[scale=.35]{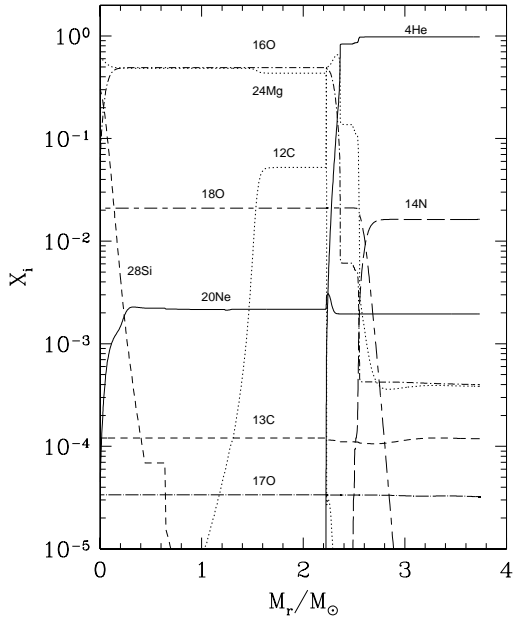}
\caption{\label{fig:abundance} Chemical structure of the primary star
 near to oxygen core exhaustion. At this stage the star is
 already devoid of hydrogen.}
\end{center} 
\end{figure}

At the time of explosion the primary is a H-free star (see
Figure~\ref{fig:abundance}) with a mass of $3.74$ $M_{\odot}$ and a
radius of $32.3$ $R_{\odot}$, in concordance with our hydrodynamical
estimations. The companion star reaches a mass of $33.7$ $M_{\odot}$, 
with luminosity and effective temperature [$\mathrm{Log}(L/L_{\odot})=
5.36$ and $\mathrm{Log}(T_{\mathrm{eff}}/\mathrm{K})=4.64$]
comparable to those corresponding to a ZAMS star of $\approx 42$
$M_{\odot}$. The companion star is thus appreciably overluminous, in
agreement with previous predictions \citep[see, e.g.,][]{dray07}.

\begin{figure}[htpb]
\begin{center}
\includegraphics[scale=.35]{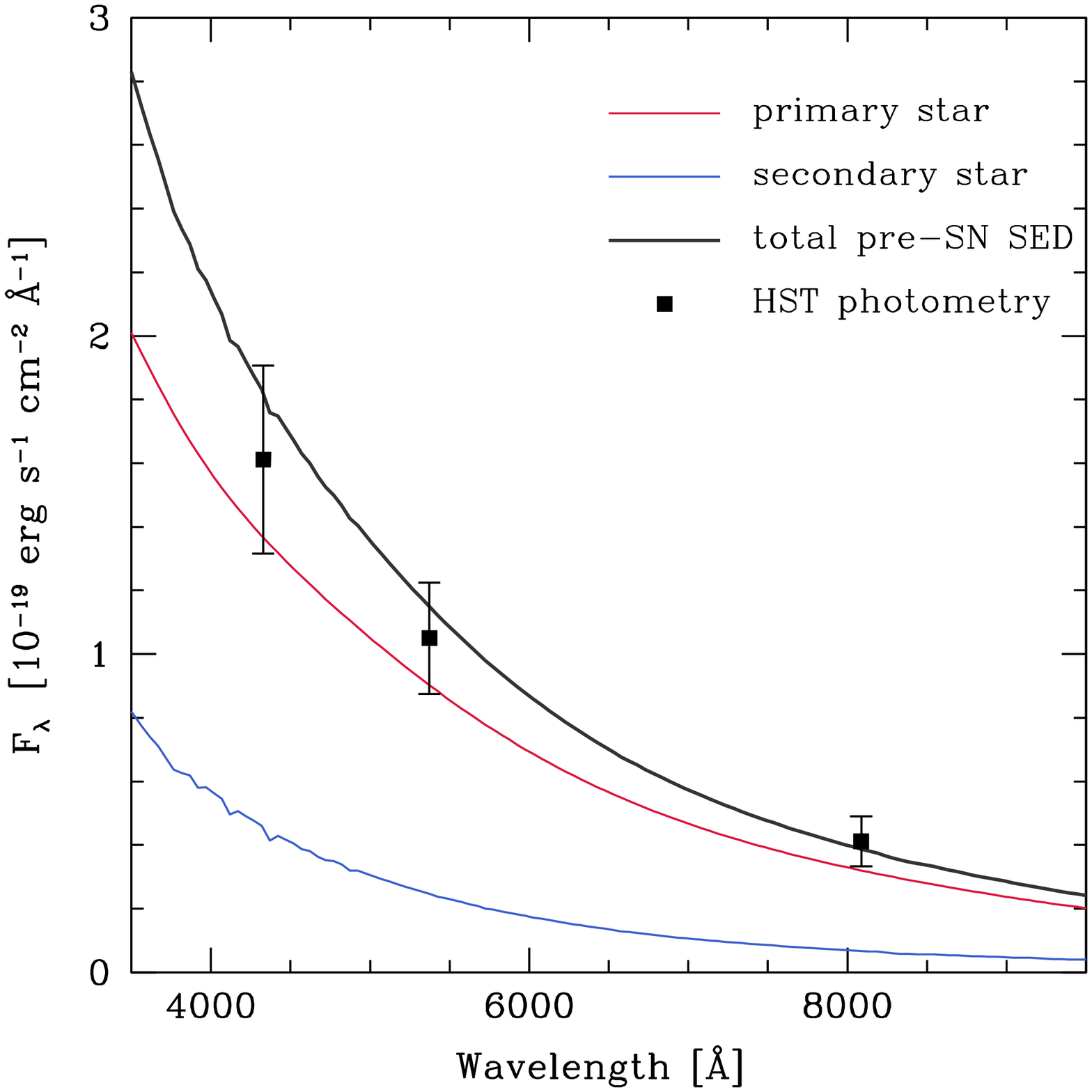}
\caption{\label{fig:SED}Predicted spectrum of the binary progenitor
  (solid black line) 
compared with HST pre-SN photometry (black squares). The binary
spectrum is the sum of a primary star approximated by a
black body (red line) and a secondary star represented by an
atmosphere model of \citet{kurucz93} (blue line). The spectra have
been extinguished assuming a standard reddening law \citep{cardelli89}
and adopting the extinction value derived in
Section~\ref{sec:data}. The HST photometry was adopted from 
\citet{cao13} and converted to specific fluxes at the approximate
effective wavelength of the F435W, F555W, and F814W bands.}
\end{center} 
\end{figure}

Figure~\ref{fig:SED} shows that the pre-SN state of the proposed
binary progenitor is compatible with the HST observations. To perform
this comparison we assumed the primary's 
spectral energy distribution (SED) is well-reproduced by a black body of the
given temperature and luminosity. This is a reasonable approximation
for a low-mass He star, as suggested by \citet{yoon12}. For the secondary
star, we adopted an atmosphere model from \citet{kurucz93} for a
main-sequence star of the corresponding effective temperature, scaled
to reproduce the required luminosity. We summed both contributions,
applied the extinction correction derived in \S~\ref{sec:data}, and
converted to observed flux adopting the distance of $25.5$ Mpc. The
synthetic photometry of the progenitor system in the three existing
bands is in agreement with the observations within the uncertainties,
with differences of less than $0.1$ mag. We note that the primary
star dominates the flux in the optical regime, so its explosion should
eventually leave a much fainter object (i.e., the 
  companion star)\footnote{Even if the secondary is bolometrically
    more luminous than the primary star, its emission in optical range
    is lower due to its hotter temperature}. The disappearance of  
  the primary star could be confirmed once the SN fades below its
  brightness. Considering the usual decline rates of SE SNe, we
  estimate to occur at about three years after 
  explosion.   

\section{PREDICTABILITY ON THE COMPANION STAR} 
\label{sec:companion}
The solution to the progenitor system presented in
Section~\ref{sec:binary} is not unique. Based on the pre-explosion
photometry and the SN observations we studied the range of
allowed binary systems with the aim of predicting the nature of the
remaining companion star. The following analysis is not intended to be an
accurate derivation but an approximation based on our calculations and
general knowledge of interacting binaries. 

The first condition for the binary scenario is that the pre-SN
structure should have a mass compatible with the light curve and
should be devoid of hydrogen. A primary star initially more massive than
$\approx 25$ $M_\odot$ would result in a helium core too massive
to reproduce the SN light curve. On the opposite
extreme, if the initial mass of the primary were $\lesssim 15$
$M_\odot$, it would be difficult to find hydrogen-free
structures. Furthermore, the range of allowed pre-explosion masses for 
the helium star (3--5 $M_\odot$) places some limits on its final
luminosity. Following \citet{yoon10,yoon12}, we consider this range to
be roughly $4.6 \lesssim \mathrm{Log}(L^{\mathrm{f}}_1/L_\odot)
\lesssim 5.0$. 

The secondary star in turn should not have a mass too close to that of the
primary to prevent it from evolving before the explosion
\citep{claeys11,benvenuto13}. If it had 
evolved, and considering the allowed range of luminosity of the primary, the
system luminosity would have become too large. This 
assures that the companion star should remain near the ZAMS. Its exact
location on the H-R diagram depends on its final mass, which in
turn is determined by the total mass of the system and the accretion
efficiency, $\beta$. In order to avoid common envelope episodes, which
is beyond the capabilities of the present code, we have additionally
required that the mass ratio of the system was not lower than
$M^{\mathrm{i}}_2/M^{\mathrm{i}}_1=0.8$\footnote{Note that this constraint 
  is not physically motivated. Therefore it may affect our
  predictability at the low end of the secondary luminosity.}. 

The above restrictions allowed us to place constraints on the final
mass---and thereby luminosity---of the companion star. Considering
approximate ranges of initial primary mass of $15 \lesssim
M^{\mathrm{i}}_1 \lesssim 25$ $M_\odot$ and initial mass ratio of $0.8
\lesssim M^{\mathrm{i}}_2/M^{\mathrm{i}}_1 \lesssim 0.95$, the final 
secondary mass results in the range of $23 \lesssim M^{\mathrm{f}}_2
\lesssim 45$ $M_\odot$. Given the uncertainties on the mass accretion
mechanism, we relaxed the condition on $\beta$ to be $0.5 \lesssim \beta
\leq 1$, and thus we found that $18 \lesssim M^{\mathrm{f}}_2 \lesssim 45$. 
Additionally, the secondary star is expected to be overluminous compared
to normal ZAMS stars of equal mass, as we found for the system
presented in Section~\ref{sec:binary}. Its 
luminosity would correspond to that of a normal ZAMS star of $\approx 20$\%
larger mass (see Figure~\ref{fig:binary}). This implies a range of
luminosities on the ZAMS of $4.6 \lesssim
\mathrm{Log}(L^{\mathrm{f}}_2/L_\odot) \lesssim 5.6$. We corroborated
that such range was not further reduced by the constraints from
the pre-explosion photometry. 

This analysis suggests that the explosion of iPTF13bvn has
left a remnant companion star of O-type characteristics. We thus predict
that this object could be recovered with future 
deep observations once the light from the SN ejecta becomes faint
enough. With the currently available instrumentation at HST, secondary
stars with $\mathrm{Log}(L/L_{\odot}) \gtrsim 5.3$ will be detectable
at $S/N \gtrsim 5$ with exposures times of the order of one hour in near-UV
and blue optical bands. 

We finally note that the HST pre-explosion photometry is in principle
also compatible with a single underluminous (by a factor of $1.5$--2) late
B- or early A-type supergiant star. Such object can be discarded as
the progenitor of a stripped-envelope SN. In order not to affect the
photometry, the actual progenitor should be at least one order of
magnitude less luminous in the optical range than the supergiant. One
possibility would be a Wolf-Rayet star, but that is ruled out by our
hydrodynamical analysis. The alternative is that the SN was produced
by the merger of two unseen compact stars. Such alternative can be
tested once the SN light fades from sight by checking whether the flux
of the remaining object has remained nearly unchanged. 

\section{CONCLUSIONS} 
\label{sec:conclusion}
Our hydrodynamical analysis of iPTF13bvn pointed to the explosion of a
low-mass helium star (of $\approx 3.5$ $M_\odot$) with a
relatively low explosion energy (of $\approx 7 \times 10^{50}$ erg)
and normal production of radioactive material 
($M_{\mathrm{Ni}}\approx 0.1$ $M_\odot$). Interestingly, from the LC
rise time we could conclude that this relatively normal event managed
to produce a quite strong \Ni\ mixing. Our conclusion about the
  low pre-SN mass is robust because of the strong constraint on the
  explosion time and it is not affected by the systematic uncertainty of
  $0.1$ dex in luminosity.
 
Our LC modeling is in contradiction with a Wolf-Rayet progenitor
for iPTF13bvn as suggested by \cite{groh13b}. In order to explain the
explosion of a low-mass helium star, we proposed the possibility of an
interacting binary progenitor. 
We showed that a system composed of 
20~$M_\odot$ + 19~$M_\odot$ stars and an initial orbital period of $4.1$
days can fully satisfy all the observational constraints
(pre-explosion mass, chemical composition and HST photometry). The
primary star is expected to dominate the flux of the progenitor 
in the optical, so as a result of the SN explosion we predict that
the flux in the observed bands will decrease significantly when the
SN fades.

We went one step beyond and studied the possible binary
configurations that could lead to compatible solutions for all 
the observational requirements with the focus on making predictions about
the putative companion star. We found that the remaining star should
necessarily be close to the ZAMS with a range of luminosities of $4.6
 \lesssim \mathrm{Log}(L^{\mathrm{f}}_2/L_\odot) \lesssim 5.6$ 
This means that the companion
star may be detected in the future with deep HST imaging in the
UV--blue range. The detection of the companion would produce the
first robust identification of a hydrogen-deficient SN progenitor
as a binary system. 

While recent evidence suggests a large fraction of massive stars belong to
interacting binary systems \citep{sana12}, it is still not clear
whether this is the main channel to produce hydrogen-free SN. The
combination  of hydrodynamical SN models and close binary evolution
calculations proves to be a powerful tool for understanding the nature
of these events in a self-consistent way.

Finally, we studied the implications on the progenitor size by modeling the
early $R$-band LC. Contrary to what might be expected from its
monotonic rise, we showed that not only compact structures (of a few
$R_\odot$) but also relatively extended envelopes ($\lesssim 150$ $R_\odot$)
are allowed with the present cadence of the observations. Our
calculations suggest that sub-night cadence is required to distinguish
among progenitor sizes in the above range. Ongoing surveys such as
KISS (Kiso Supernova Survey) or future programs like ZTF (intermediate
Palomar Transient 
Factory and Zwicky Transient Facility) will be able to provide the
necessary frequency of observations to solve this kind of problem.

\section*{Note} 
 This article took four months to be accepted due to delays by the
  referee. This accepted version is only slightly different from the
  submitted one.
\acknowledgments
We thank A.~Gal-Yam and Y.~Cao for kindly providing early-time data.
This research has been supported in part by the Grant-in-Aid for
Scientific Research of MEXT (22012003 and 23105705) and 
JSPS (23540262) and by World Premier International Research Center
Initiative, MEXT, Japan. OGB is Member of the Carrera del 
Investigador  Cient\'{\i}fico of the Comisi\'on de Investigaciones 
Cient\'{\i}ficas (CIC) of the Provincia de Buenos Aires, Argentina.
 G.~F. acknowledges financial support by Grant-in-Aid for Scientific
 Research for Young Scientists (23740175). GCA and DKS acknowledge 
partial support from DST under the India-Japan S$\&$T Cooperation programme.
Support for HK is provided by the Ministry of Economy, Development, and
Tourism's Millennium Science Initiative through grant IC12009, awarded to
The Millennium Institute of Astrophysics, MAS. HK acknowledges support
by CONICYT through FONDECYT grant 3140563.  

\clearpage

\end{document}